\documentstyle[prb,aps,psfig]{revtex}
\begin{document}
\newcommand {\be}{\begin{equation}}
\newcommand {\ee}{\end{equation}}
\newcommand {\bea}{\begin{eqnarray}}
\newcommand {\eea}{\end{eqnarray}}
\newcommand {\nn}{\nonumber}

\draft
%
%
%
%

\title{Phase Diagram and Thermodynamic Properties of the Square
Lattice of Antiferromagnetic
Spin-1/2 Triangles in $\rm \bf La_4Cu_3MoO_{12}$}

\author{Stefan Wessel and Stephan Haas}
\address{Department of Physics and Astronomy, University of Southern
California, Los Angeles, CA 90089-0484}

\date{\today}
\maketitle

\begin{abstract}
The magnetic phase diagram and the thermodynamic properties
of a square lattice containing antiferromagnetically coupled 
spin-1/2 triangles are studied. 
A Heisenberg
Hamiltonian with three strong intra-triangle exchange coupling constants
and one weak inter-triangle exchange coupling constant is proposed to 
model the interacting Cu$^{2+}$ ions of the $\rm Cu_3MoO_4$ planes 
in $\rm La_4Cu_3MoO_{12}$. 
Depending on the ratio of the intra-triangle coupling constants, various 
long-ranged magnetic phases are shown to compete at low temperatures. 
A comparison of numerical calculations of thermodynamic properties in the 
model Hamiltonian 
with recent experiments on $\rm La_4Cu_3MoO_{12}$ suggests that
the spins in this material order 
antiferromagnetically along the x-direction and 
ferromagnetically along the y-direction of the $\rm Cu_3MoO_4$ planes.
The temperature dependence of the uniform magnetic susceptibility, the
heat capacity, and the entropy are calculated and are shown to be in
qualitative agreement with the
experiments on this compound. 
\end{abstract}
\pacs{}

It is known that
the zero-point fluctuations in two-dimensional (2D) quantum
antiferromagnets are responsible for a strong renormalization
of their order parameter, causing a substantial correction
of the sublattice magnetization in long-ranged antiferromagnets
such as $\rm La_2CuO_4$ \cite{214}, or completely suppressing the
magnetic moment in short-range-ordered antiferromagnetic (AF) 
compounds such as 
the structurally frustrated Kagome lattice,
containing coupled spin-1/2 triangles.
\cite{lecheminant,mila,sindzingre}  
In contrast to systems with long-range AF order, this spin liquid
is best described within a resonant valence bond picture that 
accounts for the degeneracy of the ground state, the singlet-triplet
spin gap, and the large number of low-lying singlet states
in the thermodynamic limit\cite{lecheminant,mila,sindzingre}.

Recently, thermodynamic 
measurements on the triangular cluster compound $\rm La_4Cu_3MoO_{12}$
were reported\cite{azuma}. This system is a layered material, containing
planar $\rm Cu_3MoO_4$ square lattices of nearly isosceles
$\rm Cu^{2+}$ spin-1/2 triangles, as shown in Fig. 1.
In contrast to the Kagome antiferromagnet, $\rm La_4Cu_3MoO_{12}$ orders 
antiferromagnetically at a N\'eel temperature of 2.6K despite the 
structural frustration within the triangles. This indicates that 
- unlike the Kagome lattice - the
bipartite nature of the square lattice of triangles in this compound is
essential for the long-range ordering. 
Associated with the antiferromagnetic ordering transition, an entropy change of 
1.56 J/(K mol Cu) was observed. Furthermore, a plateau in the magnetization
above a critical field of approximately 20 T was reported.

\begin{figure}
\centerline{\psfig{figure=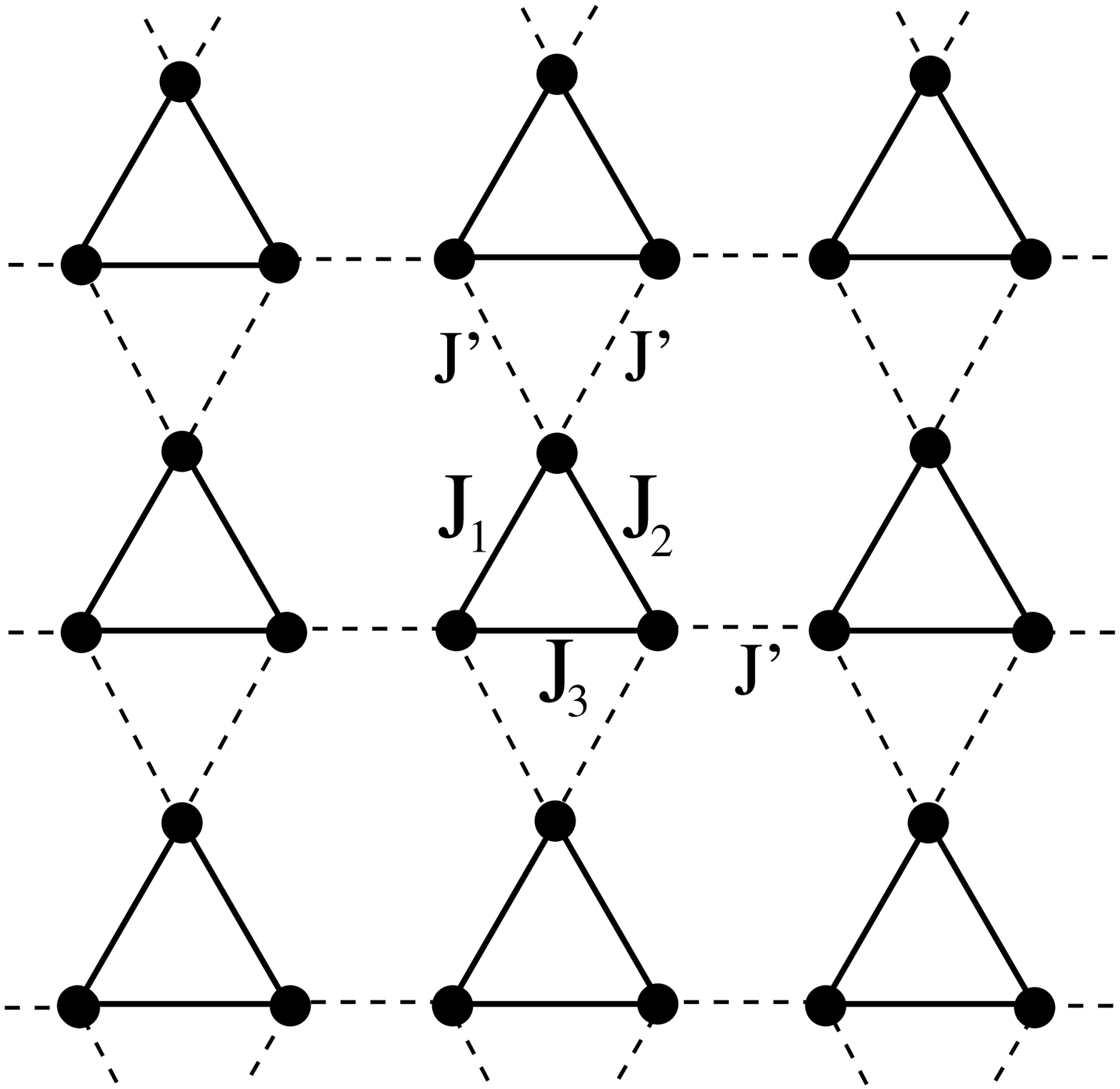,width=7cm,angle=0}}
\vspace{0.5cm}
\caption{Square lattice of Cu$^{2+}$ triangles in the
$\rm Cu_3MoO_4$ planes of $\rm La_4Cu_3MoO_{12}$.
The intra-triangle couplings $J_1,J_2,$ and $J_3$ are strong 
compared to the inter-triangle couplings $J'$.
}
\end{figure}

From the different Cu-O bond lengths within and between the triangles, 
\cite{azuma}
it can be concluded that the
intra-triangle couplings $J_1$, $J_2$, and $J_3$ are much stronger than 
the inter-triangle coupling $J'$. 
The magnetic moments of the neighboring Cu$^{2+}$ ions within
the $\rm Cu_3MoO_4$ 
planes interact via antiferromagnetic superexchange, mediated by the 
filled O$^{2-}$ p-orbitals.
This leads us to propose 
a planar model Hamiltonian of weakly coupled spin-1/2 trimers
for $\rm La_4Cu_3MoO_{12}$, given by
\begin{equation}
H=\sum_{\bf r} \left[J_1\mbox{\bf S}_{{\bf r},1}\cdot \mbox{\bf S}_{{\bf r},2}+
J_2\mbox{\bf S}_{{\bf r},2}\cdot \mbox{\bf S}_{{\bf r},3}+
J_3\mbox{\bf S}_{{\bf r},3}\cdot\mbox{\bf S}_{{\bf r},1}+ 
J'(\mbox{\bf S}_{{\bf r},3}\cdot\mbox{\bf S}_{{\bf r}+{{\bf x}},1}+ 
\mbox{\bf S}_{{\bf r},2}\cdot\mbox{\bf S}_{{\bf r}+{{\bf y}},1} + 
\mbox{\bf S}_{{\bf r},2}\cdot\mbox{\bf S}_{{\bf r}+{{\bf y}},3})\right],
\end{equation}
where  $\mbox{\bf S}_{{\bf r},j}$ represents a spin-1/2 degree of freedom 
at site $j$ of the triangle centered at position ${\bf r}$.
The underlying 2D square lattice of triangles
is spanned by the unit vectors ${{\bf x}}$ and  
${{\bf y}}$. In order to account for deviations from perfect 
isosceles triangles, 3 different intra-triangle couplings are
considered. From experience, it is rather difficult to derive the precise
exchange 
constants from first principles.\cite{bloch} An alternative approach 
is to determine them by comparing results of model calculations
on $H$
to experimental data.
However, this approach is typically
complicated by the fact that various parameter sets may fit the experimental 
results equally well.\cite{footnote1}
In the following, we will therefore first examine
the weak-coupling phase diagram of $H$.
Subsequently, the recent    
experiments on $\rm La_4Cu_3MoO_{12}$ will be analyzed, 
based on numerical calculations for the proposed
model Hamiltonian in the relevant parameter regime.

The weak-coupling ($J' \ll J_1, J_2, J_3$) zero-temperature
phase diagram of $H$ is determined 
by the dominant spin-spin correlation function of the total trimer
spins $\mbox{\bf \={S}}_{\bf r}=\mbox{\bf S}_{{\bf r}1}+\mbox{\bf S}_{{\bf r}2}+
\mbox{\bf S}_{{\bf r}3}$, defined by
$S_{\bf q} = \sum_{\bf r} \exp{(i {\bf q} \cdot {\bf r})} 
\langle \mbox{\bf \={S}}_{\bf 0} \cdot \mbox{\bf \={S}}_{\bf r} \rangle$. We have calculated this quantity
by exact numerical diagonalization of finite clusters with 
periodic boundary conditions  
for small inter-triangle coupling strengths $J'$.\cite{footnote2}
It is found that the wave vector
of the dominant spin-spin correlation function
strongly depends on the relative strengths of the 
intra-triangle couplings $J_1$, $J_2$, and $J_3$.
The resulting zero-temperature phase diagrams for square lattices
of weakly coupled triangles are shown in Fig. 2 with (a) $J'=0.01 J_3$,
and (b) $J'=0.1 J_3$, where the various
phases are labeled by their magnetic ordering 
wave vectors ${\bf Q} = (q_x,q_y)$.

\begin{figure}
\centerline{\psfig{figure=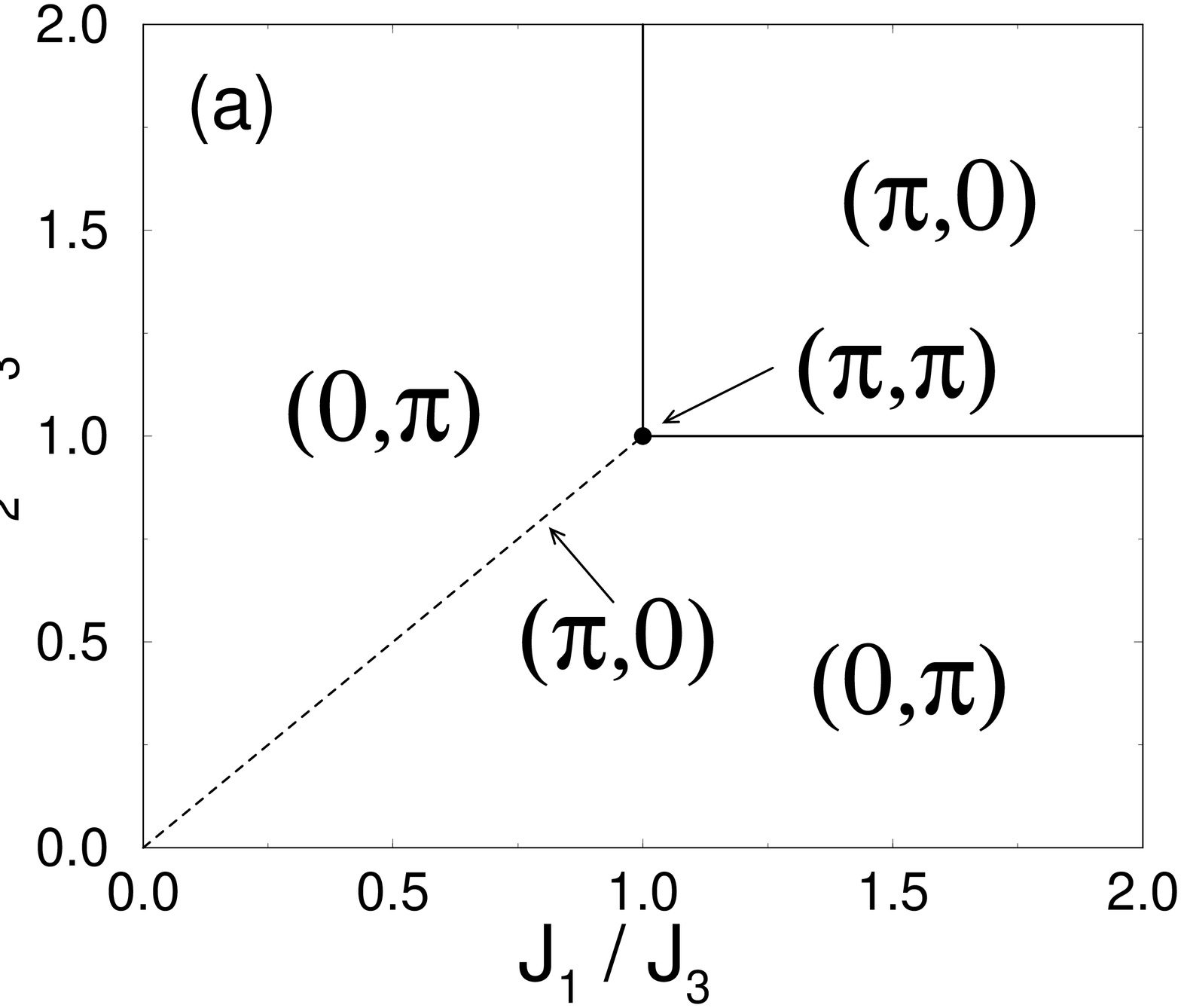,width=5cm,height=5cm,angle=270}\hspace{1.0cm}\psfig{figure=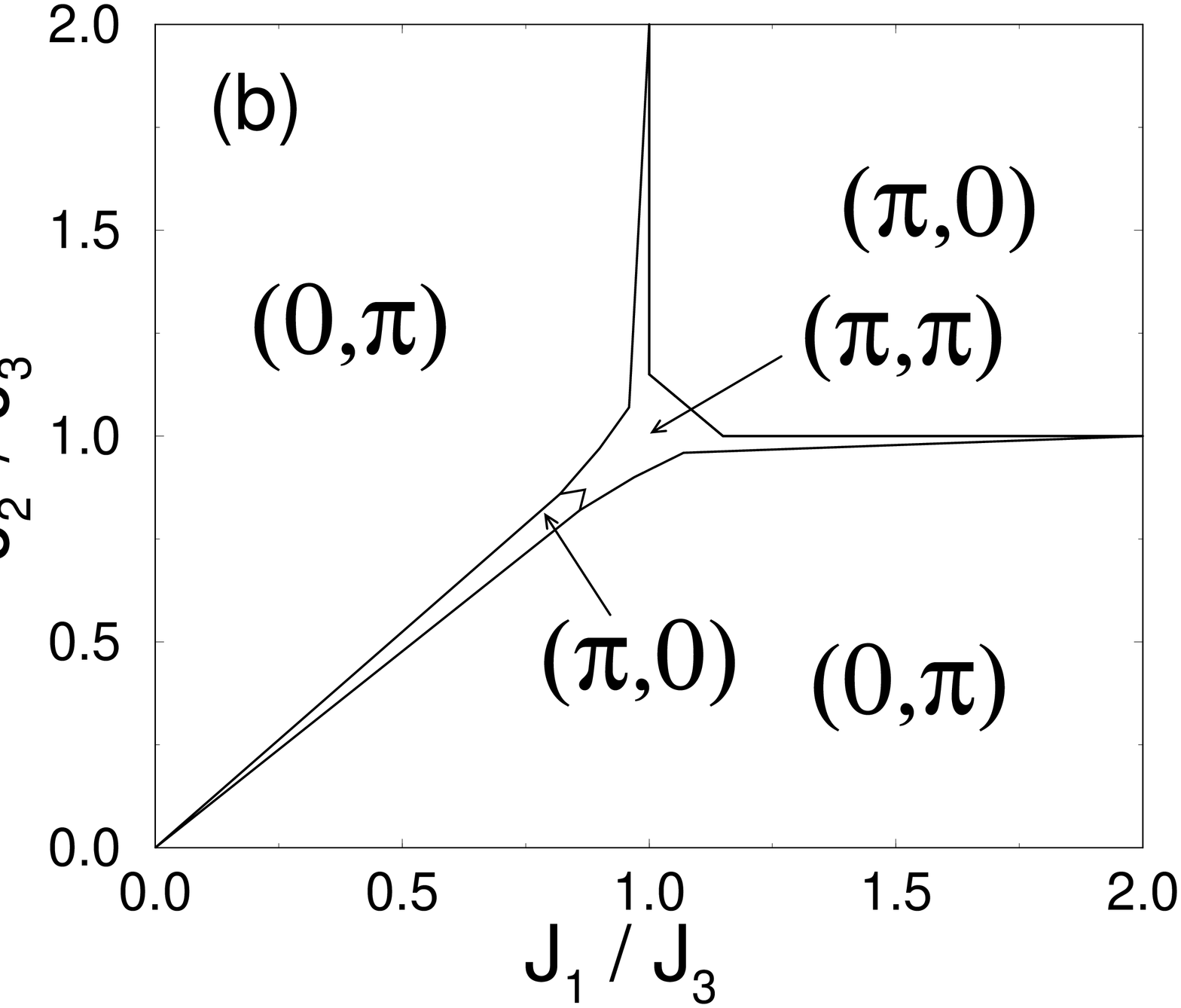,width=5cm,height=5cm,angle=270}}
\vspace{0.5cm}\caption{Zero-temperature 
magnetic phase diagram of the AF
spin-1/2 trimer square lattice with a weak inter-triangle coupling
(a) $J'=0.01 J_3$ and (b) $J'=0.1 J_3$. 
}
\end{figure}

Since the model Hamiltonian $H$ 
is invariant under $J_1 \leftrightarrow J_2$, these phase diagrams 
have reflection symmetry about the diagonal axis $J_1 = J_2$. 
All exchange coupling constants have been scaled by the
intra-triangle coupling constant along the x-direction, $J_3$.
In Fig 2(a), the phase diagram is shown for the
ultra-weak-coupling limit with a small
inter-triangle coupling $J' = 0.01 J_3$.
At the isotropic point $J_1=J_2=J_3$, the system orders
antiferromagnetically in both planar
directions, resulting in an ordering wave vector ${\bf Q} = (\pi, \pi)$.
For small $J_3$ ($J_3 < J_1,J_2$), the system decouples into
spin-1/2 AF chains along the x-direction with weak residual
ferromagnetic (FM) couplings along the y-direction,
leading to ${\bf Q} = (\pi,0)$. In contrast, in the regimes $J_1 \gg J_2, J_3$ 
and $J_2 \gg J_1,J_3$, the model describes effective spin-1/2
AF chains along the y-axis with weak residual FM coupling along the 
x-direction, and thus ${\bf Q} = (0,\pi)$. The symmetry line 
$J_1=J_2$ with ${\bf Q} = (\pi,0)$ is found to
separate these two ${\bf Q} = (0,\pi)$ phases.\cite{footnote3}

As the inter-triangle coupling $J'$ is increased (Fig. 2(b)), a finite
region with ${\bf Q} = (\pi, \pi)$ opens up, centered around the isotropic
point $J_1=J_2=J_3$. Furthermore, the ${\bf Q} = (\pi,0)$ regime is found
to extend away from the diagonal $J_1=J_2$ line.
It has been proposed that the AF ordered low-temperature
phase of $\rm La_4Cu_3MoO_{12}$ may fall into the 
${\bf Q} = (\pi, \pi)$ regime.\cite{azuma}
However, as we will discuss in the following,
a numerical analysis of the thermodynamic properties of $H$ - in particular the
magnetic entropy - suggests that this compound should
be placed in the regime of weak $J_3$ with  $ {\bf Q} = (\pi,0)$ 
rather than in the ${\bf Q} = (\pi, \pi)$ regime close to the isotropic
point.  

When analyzing the thermodynamic response functions for the square lattice of 
weakly coupled AF spin-1/2 triangles, three temperature regimes can be 
identified. At high 
temperatures $T>J_1,J_2,J_3$, the system is a paramagnetic ensemble of 
thermally decoupled spins. Upon lowering the temperature into the region 
$J'<T<J_1,J_2,J_3$, the intra-triangle couplings become relevant, and   
spin doublets are formed on each triangle. The properties of the system in this 
temperature interval are thus well characterized by a thermodynamic
ensemble of decoupled trimers with 
effective spin-1/2 degrees of freedom. Below the ordering transition
temperature $T_N \approx J'$ the small inter-triangle couplings correlate
these spin-1/2 trimers, leading to long-range-ordered magnetic
phases with characteristic $\bf Q$ vectors that depend on the ratio of the 
intra-triangle coupling constants, as discussed above. 
We have determined the
temperature dependence of the uniform magnetic susceptibility, 
the heat capacity, and the entropy from full numerical diagonalizations 
of $H$ on finite clusters with periodic boundary conditions.
Results representative for the 
the $ {\bf Q} = (\pi,0)$ phase with $(J_1,J_2,J_3) = (1.5,1.5,1.0)$, 
$ {\bf Q} = (0,\pi)$ phase with $(J_1,J_2,J_3) = (0.5,1.0,1.0)$, and the 
${\bf Q} = (\pi, \pi)$ phase with $(J_1,J_2,J_3) = (1.0,1.0,1.0)$
are shown in Fig. 3. 

\begin{figure}
\centerline{\psfig{figure=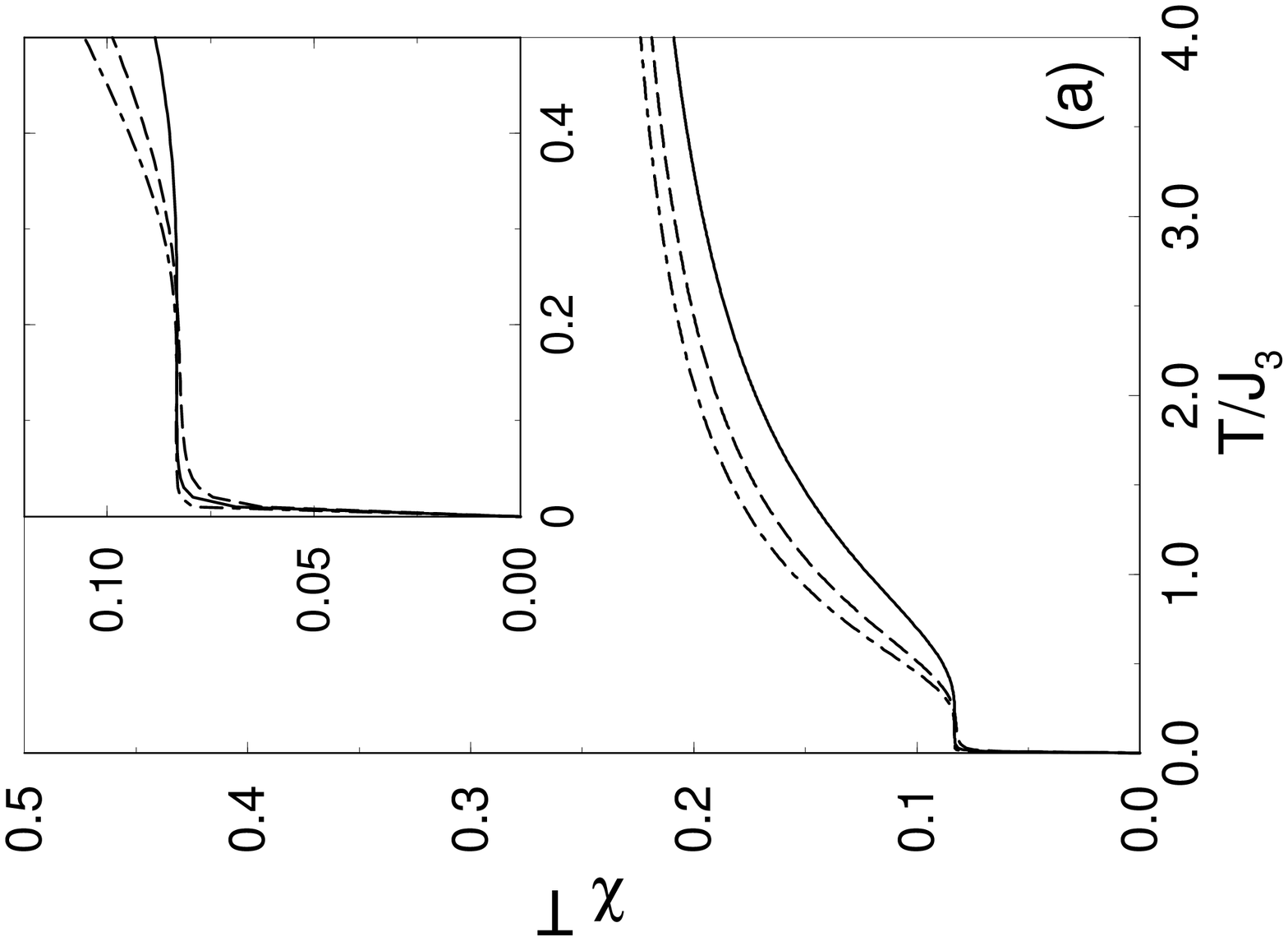,width=5cm,angle=270}
\psfig{figure=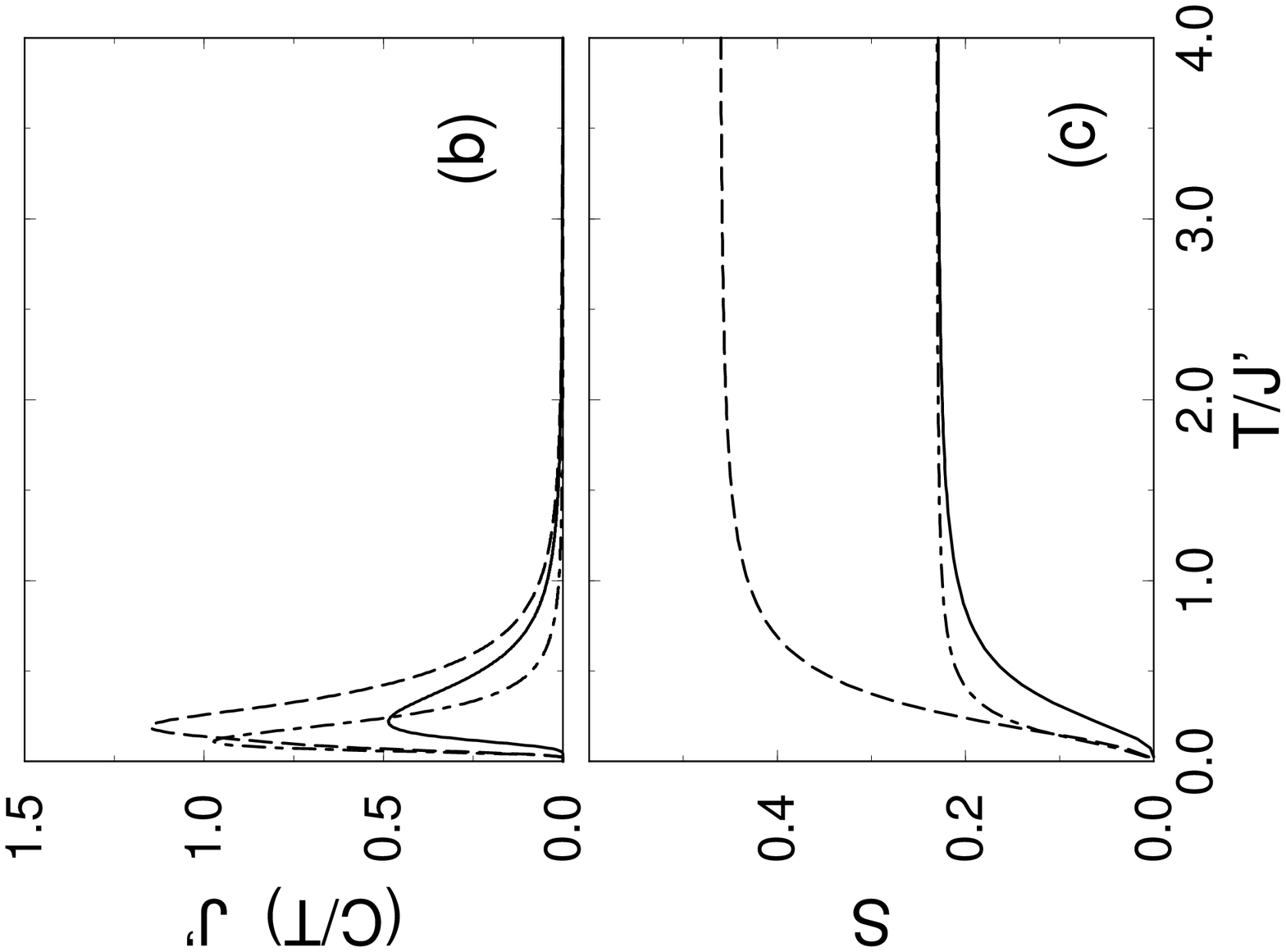,width=5cm,angle=270}}
\vspace{0.5cm}\caption{Thermodynamic response in
the square lattice of AF spin-1/2 triangles.
(a) Temperature-dependent uniform susceptibility
in the $ {\bf Q} = (\pi,0)$ regime (solid line), the 
$ {\bf Q} = (0,\pi)$ regime (dot-dashed line), and the 
$ {\bf Q} = (\pi,\pi)$ regime (dashed line).
The inset shows the intermediate-temperature plateaus of $\chi T$ at
$\frac{1}{12}$. (b) Temperature dependence of the specific heat, and
(c) the magnetic entropy per site. 
}
\end{figure}

The three distinct temperature regimes are clearly observed in 
the uniform susceptibilities, shown in Fig. 3(a). 
At high temperatures the spins are effectively uncorrelated
with a Curie constant 
$\lim_{T \rightarrow \infty} [\chi T ] = C=\frac{1}{4}$. 
At intermediate temperatures there is a plateau in $\chi T$
with a reduced effective Curie constant $C'=\frac{1}{12}$,
associated with the paramagnetic
response of uncorrelated spin-1/2 trimers, containing three
spins each. Finally, at low temperatures of order $J'$ these 
trimers condense into magnetically ordered phases, leading to
an additional reduction in $\chi T$. These trends in the temperature 
dependence of the uniform susceptibilities are similar in all 
three parameter regimes. Furthermore, the shape of $\chi (T)$
is found to be in good qualitative agreement with measurements 
on $\rm La_4Cu_3MoO_{12}$ (see Fig. 2 of Ref. \cite{azuma}),
which show an onset of magnetic ordering at
2.6 K, two orders of magnitude smaller than the observed threshold
temperature for trimer formation at approximately 250 K.

In Figs. 3 (b) and (c) the specific
heat and the magnetic entropy per site are shown. 
The onset of the intermediate
regime in which intra-triangle correlations become relevant is marked
by the maxima in the specific heat divided by the temperature, $C/T$,
consistent with the onsets of the plateaus in $\chi T$, seen in Fig. 3(a). 
For the $ {\bf Q} = (\pi,0)$ and the $ {\bf Q} = (0,\pi)$ regimes, 
the entropies per site exhibit plateaus
at $S = \frac{1}{3} \ln 2$, rising up to $\ln 2$ at high temperatures
(not shown in the figure). 
Hence, only 1/3 of the magnetic entropy is released in the intermediate
temperature regime, corresponding to 
an ensemble of N/3 trimer doublets with a decreased paramagnetic
moment of spin-1/2 per trimer.
In contrast, the entropy for the isotropic coupling case with 
$ {\bf Q} = (\pi , \pi)$ shows a larger plateau value of
$\frac{1}{3} \ln 4$, associated with the degeneracy of two Kramers doublets
in this regime. In fact, the same plateau value is  found over the whole ${\bf Q}=(\pi,\pi)$ regime centered around the isotropic coupling point.    At very high temperatures the entropy saturates at 
the fully paramagnetic value of $\ln 2$ in all cases. A comparison
with the experiments on $\rm \ La_4Cu_3MoO_{12}$ (Fig. 3 in Ref.
\cite{azuma}) clearly favors the
smaller intermediate-temperature
plateau value of $\frac{1}{3} \ln 2$ over $\frac{1}{3} \ln 4$,
suggesting that the 
low-temperature magnetic order in this material is characterized by 
$ {\bf Q} = (\pi,0)$ or $ {\bf Q} = (0,\pi)$ rather than by 
$ {\bf Q} = (\pi , \pi)$. 

Examining the intra-triangle bond lengths and bond angles in
the $\rm Cu_3MoO_4$ planes of $\rm  La_4Cu_3MoO_{12}$ (table 1
of Ref. \cite{azuma}), it appears likely that $J_3$ is smaller than 
both $J_1$ and $J_2$.\cite{bloch,footnote4} This suggests that
the magnetically ordered low-temperature phase
in this compound is part of the $ {\bf Q} = (\pi,0)$ regime rather
than the $ {\bf Q} = (0,\pi)$ regime. Neutron scattering
studies should help to verify this conjecture.

In summary, we have studied the magnetic phase diagram and the thermodynamic 
response of a square lattice containing weakly coupled AF spin-1/2 triangles. 
A Heisenberg
Hamiltonian with three strong intra-triangle exchange coupling constants
and a weak inter-triangle exchange coupling constant is proposed to 
model the interacting Cu$^{2+}$ ions of the $\rm Cu_3MoO_4$ planes 
in $\rm La_4Cu_3MoO_{12}$. 
Depending on the ratio of the intra-triangle coupling constants, various 
long-ranged magnetic phases are possible at low temperatures. 
A comparison of numerical calculations of the thermodynamic properties in the 
proposed model Hamiltonian 
with recent experiments on $\rm La_4Cu_3MoO_{12}$ leads us to suggest that
the magnetically ordered phase in this compound belongs to the
$ {\bf Q} = (\pi,0)$ regime rather than the $ {\bf Q} = (0,\pi)$ or
the  $ {\bf Q} = (\pi ,\pi)$
regimes of the model. The ordering wave vector could be determined
in a neutron scattering study, giving conclusive information 
regarding the nature of the magnetically ordered phase in this material.

We wish to thank M. Azuma, A. Honecker, B. Normand, and M. Takano 
for useful discussions.
The hospitality of the Los Alamos National Laboratory 
and the financial support of the Zumberge foundation are 
acknowledged.


\begin{references}
\bibitem{footnote2} We have studied lattices with 6, 12, and 18 spins.
\end{references}
\end{document}